\documentclass[12pt,preprint]{aastex}
\usepackage{longtable}

\newcommand{\hi}{H~{\small{I}}}

\shorttitle{Nearby Clumpy, Gas Rich SFGs}
\shortauthors{Garland et al.}

\begin{document}

\title{Nearby Clumpy, Gas Rich, Star Forming Galaxies: Local Analogs of High Redshift Clumpy Galaxies}

\author{C. A. Garland\altaffilmark{1}}
\affil{Natural Sciences Department, Jeffords Science Center, Castleton State 
College, Castleton, VT 05735, USA}
\altaffiltext{1}{Also Department of Astrophysics, American Museum of Natural History, Central Park West at 79$^{th}$ Street, New York, NY 10024, USA}
\email{catherine.garland@castleton.edu}

\author{D. J. Pisano}
\affil{Department of Physics and Astronomy, West Virginia University, 135 Willey Street, P.O.~Box~6315, Morgantown, WV 26506, USA}
\email{djpisano@mail.wvu.edu}

\author{M.-M. Mac Low\altaffilmark{2}}
\affil{Department of Astrophysics, American Museum of Natural History, Central Park West at 79$^{th}$ Street, New York, NY 10024, USA}
\altaffiltext{2}{Also Zentrum f\"ur Astronomie der Universit\"at Heidelberg, Institut f\"ur Theoretische Astrophysik, Albert-Ueberle-Str. 2, 69120 Heidelberg, Germany}
\email{mordecai@amnh.org }

\author{K. Kreckel}
\affil{Max-Planck-Institut f\"ur Astronomie, K\"onigstuhl 17, D-69117 Heidelberg, Germany}
\email{kreckel@mpia.de}

\author{K. Rabidoux}
\affil{Department of Physics and Astronomy, West Virginia University, 135 Willey Street, P.O.~Box~6315, Morgantown, WV 26506, USA}
\email{krabidou@mix.wvu.edu}

\author{R. Guzm\'an}
\affil{Department of Astronomy, University of Florida, 211 Bryant 
Space Science Center, P.O.~Box~112055, Gainesville, FL 32611, USA}
\email{guzman@astro.ufl.edu}

\begin{abstract}
Luminous compact blue galaxies (LCBGs) have enhanced star formation rates and compact morphologies.  We combine Sloan Digital Sky Survey data with \hi~data of 29 LCBGs at redshift z~$\sim$~0 to understand their nature. We find that local LCBGs have high atomic gas fractions ($\sim$50\%) and star formation rates per stellar mass consistent with some high redshift star forming galaxies.  Many local LCBGs also have clumpy morphologies, with clumps distributed across their disks.  Although rare, these galaxies appear to be similar to the clumpy star forming galaxies commonly observed at z~$\sim$~1$-$3.  Local LCBGs separate into three groups: 1. Interacting galaxies ($\sim$20\%);  2. Clumpy spirals ($\sim$40\%); 3. Non-clumpy, non-spirals with regular shapes and smaller effective radii and stellar masses ($\sim$40\%).  It seems that the method of building up a high gas fraction, which then triggers star formation, is not the same for all local LCBGs.  This may lead to a dichotomy in galaxy characteristics.  We consider possible gas delivery scenarios and suggest that clumpy spirals, preferentially located in clusters and with companions, are smoothly accreting gas from tidally disrupted companions and/or intracluster gas enriched by stripped satellites. Conversely, as non-clumpy galaxies are preferentially located in the field and tend to be isolated, we suggest clumpy, cold streams, which destroy galaxy disks and prevent clump formation, as a likely gas delivery mechanism for these systems.  Other possibilities include smooth cold streams, a series of minor mergers, or major interactions.
\end{abstract}


\section{INTRODUCTION}
\label{introduction}
Irregularly shaped, clumpy star forming galaxies (SFGs) that do not fall anywhere on the Hubble sequence appear frequently at redshifts z~$\sim$1$-$4 (e.g. \citealt{Cowie1995, Elmegreen2007, Guo2014} and review by \citealt{Shapley2011}).  Galaxies at z~$\lesssim$~1 with similar morphologies and enhanced star formation have been identified, but become less common with decreasing redshift \citep{Nair2010, Puech2010, Elmegreen2013, Murata2014}.  We find that close visual examination of local luminous compact blue galaxies (LCBGs) reveals them to have a high fraction of asymmetric, clumpy morphologies.  This leads us to use multi-wavelength observations to determine whether these could be local analogs to high-redshift SFGs.

Observations (e.g. \citealt{Daddi2010, Tacconi2013}) and simulations (e.g. \citealt{Noguchi1998, Overzier2008, Agertz2009, Dekel2009, Tonnesen2011, Mandelker2013}) suggest these clumpy, irregularly shaped high redshift SFGs form in two ways: 1. Due to interactions with other galaxies or cluster potentials, or 2. Due to violent gravitational instabilities and disk fragmentation driven by ongoing gas accretion, leading to high gas fractions.  The infalling gas may be delivered via cold flows or accreted directly from companions or from an intracluster medium polluted by stripped satellites.  Theoretically, all of these scenarios can give rise to the large clumps of enhanced star formation observed in such galaxies.  

From morphology alone it is difficult to distinguish between formation scenarios driven by external interactions such as galaxy-galaxy or galaxy-cluster interactions, and internal causes such as high gas fractions due to ongoing accretion (e.g. \citealt{Overzier2010}).  One important clue has been argued to be whether or not such galaxies are rotating disks.  Molecular line observations of clumpy SFGs at z~$\sim$~1$-$3 indicate most are rotating disks, not merging systems \citep{Daddi2010, Tacconi2013}. However, \citet{Forster2009} and \cite{Law2009} find evidence from integral field spectroscopy of clumpy SFGs at similar redshifts that some sources are clearly merging or interacting systems.  \cite{Law2009} caution that their strongest merger candidate also shows rotational structure, so there is not necessarily a disk versus merger bimodality.  Simulations suggest that accretion via cold streams is the dominant trigger of star formation at high redshift so that, while major merger-induced star formation is spectacular, such starbursts are uncommon \citep{Dekel2009Nature, Faucher2011}.  
  
The picture is not any clearer at z~$<$~1.  \citet{Puech2010} studied clumpy galaxies at z~$\sim$~0.6 and concluded, based on the complex kinematics of nearly half the sample, that interactions and not cold streams probably dominate at that redshift.  \citet{Luo2014} found that while $\sim$50\% of  z~$\sim$~0 vigorously starbursting galaxies are evident mergers, this is true for fewer than 20\% of local, less vigorously star forming galaxies.  \cite{Bournaud2012} used optical emission line observations to compare clumpy and smooth SFGs at z~$\sim$~0.7 and concluded that the clumpy SFGs exhibited properties of gas-rich unstable disks and not mergers.  While some simulations (e.g. \citealt{Dekel2006, Kraljic2012}) suggest that mergers and cold accretion should persist down to lower redshifts for low mass systems in low density environments, newer simulations by \citet{Nelson2014} find that cold accretion may be highly suppressed at low redshift.  At both high and low redshift there have been a range of criteria used to select clumpy SFGs.  For example, sources may be selected using optical or UV restframe wavelengths and the definition of clumpy varies from one author to another.  This makes quantitative comparisons between data sets challenging.

While z~$\sim$~0 irregularly shaped, clumpy galaxies were observed as early as the 1970's \citep{Casini1976}, catalogued \citep{Nair2010}, and had clump characteristics measured \citep{Elmegreen2013}, no prior study of these rare analogs to high redshift SFGs simultaneously measured gas content, stellar masses, star formation rates (SFRs) and metallicities.  We make such measurements, as well as determine if the galaxies have companions or lie within clusters.  This evidence allows us to examine whether their morphologies and enhanced star formation rates are caused primarily by galaxy-galaxy and galaxy-cluster interactions, by accretion of gas from satellites, or by cold streams.

We present results of our study of nearby (D~$<$~200~Mpc) LCBGs, which we selected to match the observed properties of 0.4~$\lesssim$~z~$\lesssim$~0.6 LCBGs. We have studied these bright, compact SFGs, many with clumpy appearances, at a range of wavelengths (e.g. \citealt{LCBGI, LCBGII, Perez2011}).  In this paper, we combine Sloan Digital Sky Survey (SDSS, \citealt{SDSS}) data, including environmental density measurements, with single-dish \hi~observations to probe the cause of this local, enhanced star formation and determine if these sources are local analogs to high-redshift SFGs.  Throughout the paper we apply a cosmology with $\Omega_m=0.27$, $\Omega_\Lambda=0.73$, and $H_0=70$~km~s$^{-1}$ Mpc$^{-1}$.

\section{OBSERVATIONAL RESULTS}
\label{local_SFGs}
\subsection{Sample}
\label{sample}
LCBGs are defined as having $M_B\leq-18.5$, a $B-$band surface brightness within the effective radius, SBe, brighter than 21 $B-$mag arcsec$^{-2}$, and $B-V\leq0.6$.  \citet{Werk2004} developed these criteria to select local galaxies based on the properties of LCBGs at z~$\sim$~0.4~$-$~0.6 where they are common. LCBGs are not a separate class of galaxy, but a small, extreme subset of the blue sequence.  They have high SFRs relative to their stellar masses, near solar metallicities, and include spiral, elliptical, asymmetric and peculiar morphologies with underlying older stellar populations.  Their $\sim$10$^9$~M$_\sun$ stellar masses, luminosities and high metallicities distinguish them from Blue Compact Dwarf galaxies.  LCBGs become increasingly rarer with decreasing redshift (see \citealt{Guzman1997} and \citealt{LCBGI} for a comprehensive description of LCBGs).
We began studying local LCBGs to try to understand the mechanisms that trigger and quench star formation, but, as we proceeded with this study, we found that many properties (detailed in the following sections) resemble those of high-redshift SFGs.

From the seventh SDSS data release (DR7, \citealt{SDSSDR7}), we used the defining magnitude, surface brightness and color criteria to identify $\sim$2300 local (D~$<$~200~Mpc) LCBGs with known redshifts.  We have been able to acquire, either ourselves or via archives, single-dish \hi~observations of 163 of these galaxies.  In this paper, we focus on all of these LCBGs that are located within 76~Mpc (z~=~0.0175) and with available H$\alpha$ SFRs, stellar masses and metallicities (see Section \ref{MPA-JHU}).  The distance limit was chosen to ensure the detection of both clumps and nearby companions.  Given the SDSS point spread function of 1.4$''$, clumps with a diameter of at least 0.5~kpc should still be resolved at this distance.  The SDSS magnitude limit, $m_r$(Petrosian)~$<$~17.77, means that a companion as faint and blue as the Small Magellanic Cloud would still be visible at this distance as well.  This yields a sub-sample of 29 galaxies; the larger sample of local LCBGs will be fully presented in future work.

\subsection{Optical Properties}
\label{optical_properties}
\subsubsection{Morphology}
\label{morphology}
\begin{figure}[ht]
\epsscale{1}
\plotone{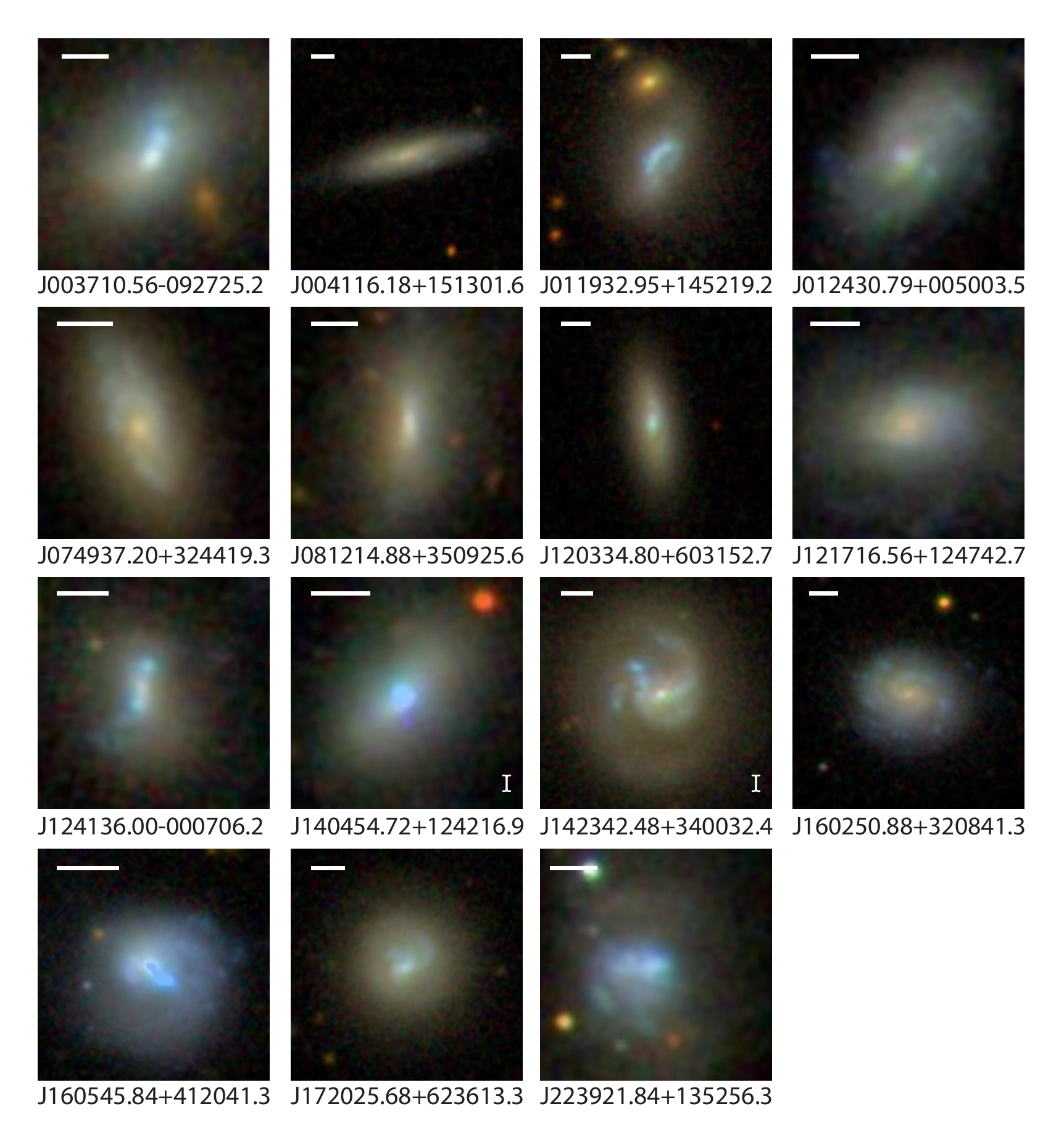}
\caption{Color-combined SDSS images of the 15 local LCBGs classified as not clumpy.  A range of morphological properties is present in our sample.  The intensity has been adjusted to optimize the appearance of each galaxy.  All sources are near enough that 0.5 kpc or larger sized clumps should be resolved. SDSS source names are included below each galaxy. A 2~kpc bar is included in the upper left corner of each image.  Galaxies designated as likely interacting are indicated with an I.}
\label{fig:LCBG_panel_not_clumpy}
\end{figure}

\begin{figure}[ht]
\epsscale{1}
\plotone{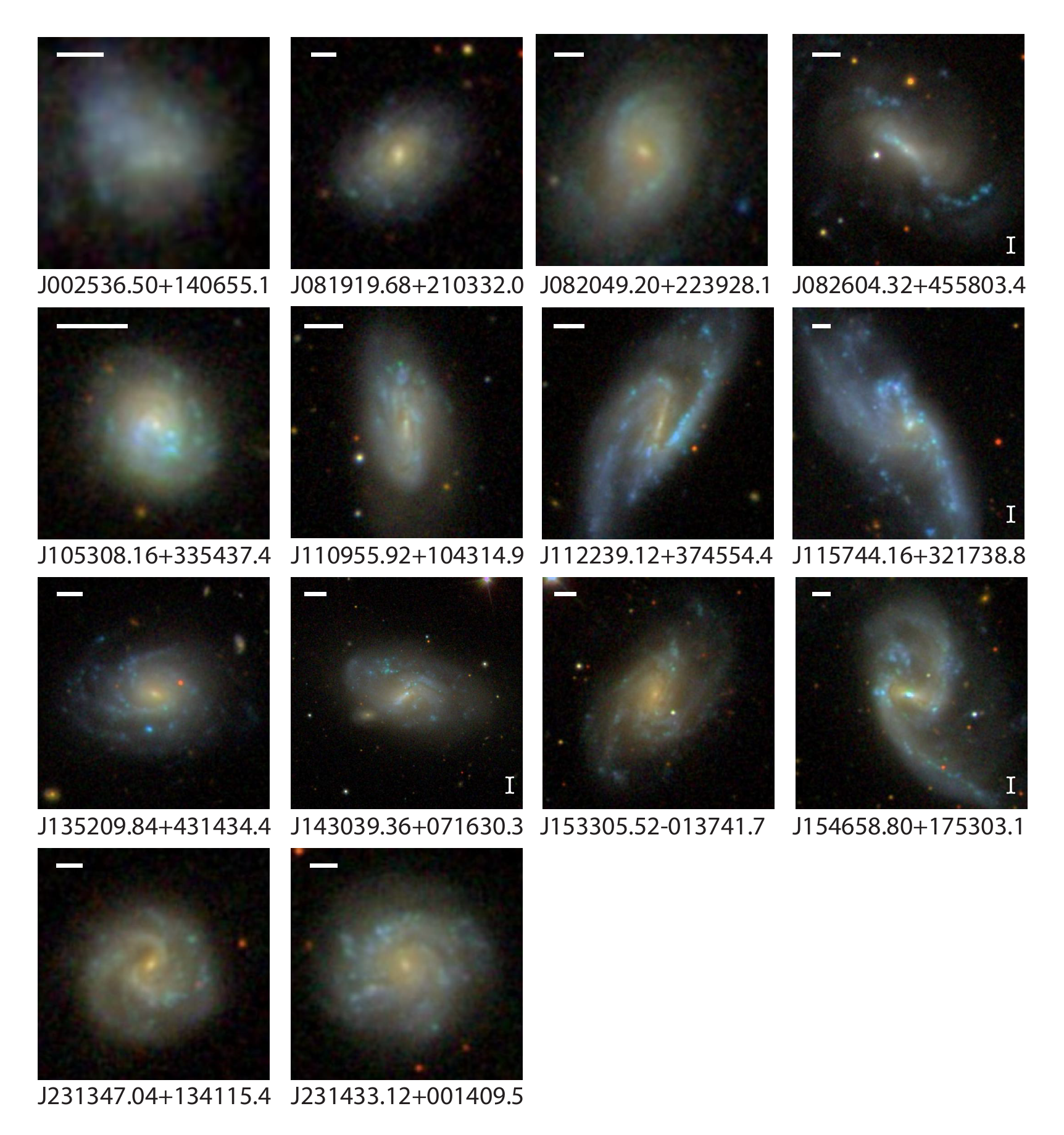}
\caption{Color-combined SDSS images of the 14 local LCBGs classified as clumpy.  A range of morphological properties is present in our sample.  The intensity has been adjusted to optimize the appearance of each galaxy.  All sources are near enough that 0.5 kpc or larger sized clumps should be resolved. SDSS source names are included below each galaxy.  A 2~kpc bar is included in the upper left corner of each image.  Galaxies designated as likely interacting are indicated with an I.}
\label{fig:LCBG_panel_clumpy}
\end{figure}

The SDSS color-combined images of our sample of LCBGs in Figures~\ref{fig:LCBG_panel_not_clumpy} and \ref{fig:LCBG_panel_clumpy} illustrate the range of morphologies present, including spiral, oval and irregularly shaped galaxies.  These images, available through the SDSS Navigate Tool\footnote{http://cas.sdss.org/dr7/en/tools/chart/navi.asp}, are made by combining corrected {\it{g}}, {\it{r}}, and {\it{i}} frames and adjusting the intensity to optimize the appearance of each galaxy \citep{SDSSEDR}. Visually, many LCBGs resemble the irregularly shaped, clumpy galaxies found at z~$\sim$~0.2$-$3 (e.g. \citealt{Elmegreen2009, Murata2014}).  This resemblance motivated us to consider whether local LCBGs might be low redshift analogs of the high redshift clumpy galaxies.  Therefore, we began to quantify the similarity between our local LCBG sample and the high redshift objects.
 
We visually identified galaxies as clumpy or not based on the number of clumps seen in the SDSS color-combined images.  Any galaxy with three or more clumps (not including the nucleus) was classified as clumpy regardless of overall morphology.  Two of the authors independently classified the galaxies with no differences in whether or not they were designated clumpy.  The classification was done using large images displayed on computer screens.  Figure~\ref{fig:LCBG_panel_not_clumpy} includes all sources we classified as not clumpy while Figure~\ref{fig:LCBG_panel_clumpy} shows all clumpy sources.  A few galaxies classified as not clumpy may appear to have three or more non-nuclear clumps in the smaller images presented in Figure~\ref{fig:LCBG_panel_not_clumpy}. However, when larger versions of these images are examined, these galaxies appear ``patchy'' rather than having obvious clumps.  In addition, some objects that look like clumps are classified as stars by SDSS and so were not counted.  A variety of methods, including visual (e.g. \citealt{Elmegreen2004, Puech2010}) and computational (e.g. \citealt{Bournaud2012, Murata2014}), have been used by others to identify clumps in galaxies. Visual classification has been found to be as good or better than computational methods \citep{Bournaud2012, Nair2010}.  Note that in other studies of clumpy galaxies, a variety of other selection criteria have been used as well.  These have included limiting galaxies to non-interacting or ``disky'' systems (e.g. \citealt{Bournaud2012}), setting a relative clump brightness minimum (e.g. \citealt{Murata2014}) or rejecting any sources with exponential disks (e.g. \citealt{Puech2010}).  We find that 48\% $\pm$ 13\% of our local LCBG sample is clumpy with clump diameters $\sim$1~kpc.  (The associated uncertainty is due to Poisson statistics.)

The identification of clumps is limited by the resolution of the respective instrumentation used.  For example, because of the SDSS PSF, for the more distant galaxies in our sample we can only state that they do not have clumps larger than $\sim$0.5~kpc.  Our rough approximation of clump sizes in local LCBGs was made using the visual extent of the clumps, the SDSS pixel scale and the distance to each galaxy.  The scope of this paper does not include a rigorous analysis of clump properties and we do not report rigorously measured clump diameters.  As some nearer galaxies have clumps visually estimated to be smaller than 0.5~kpc, it is possible some galaxies we classified as non-clumpy do have three or more clumps that are not resolved.  When examining our larger sample of local LCBGS we did find that the estimated clump sizes in these galaxies decreased as the instrumental resolution increased.  We also found that one LCBG which we had classified as non-clumpy had multiple clumps in an archival Hubble Space Telescope (HST) image with much higher resolution.   At higher redshifts, most imaging of clumpy galaxies has been done with the HST (PSF~$\sim$~0.1$''$).  Therefore, at z~$\sim$~0.02 (our maximum distance) SDSS can achieve higher resolution than the HST can at z~$\sim$~1$-$3,  meaning galaxies identified as clumpy in the local Universe may not be identified as clumpy if placed at higher redshifts.  Indeed, surveys of high-redshift lensed galaxies, which are not resolution-limited, measure clump sizes down to $\sim$300~pc \citep{Jones2010}, much smaller than the typical size of $\sim$1~kpc in non-lensed high-redshift galaxies.  However, our local sample does include some genuinely non-clumpy galaxies$-$sources that are near enough to resolve clumps of $\sim$200$-$300~pc but that are classified as non-clumpy.  

\subsubsection{Star Formation Rates, Stellar Masses, Metallicities and Effective Radii}
\label{MPA-JHU}
We retrieved H$\alpha$ SFRs, stellar masses (M$_\star$) and oxygen abundances (which we interpret as metallicity indicators) from the MPA-JHU~SDSS~DR7 Value Added Data Catalog\footnote{http://www.mpa-garching.mpg.de/SDSS/DR7/}.  The SFRs in this database were calculated using emission line luminosities and corrected for the 3$''$ fiber size following \citet{Brinchmann2004}.  The stellar masses were calculated by comparing stellar absorption line indices with broad-band photometry based on the work of of \cite{Kauffmann2003} and \citet{Salim2007}.  Finally, the gas phase metallicities were calculated from optical nebular emission lines using a Bayesian technique based on \citet{Tremonti2004}.

We summarize the properties of our local sub-sample of LCBGs in Table~\ref{table:properties}.  The SFR and M$_\star$ values are lower than those measured in higher redshift clumpy SFGs; the lower masses are expected as a natural consequence of cosmic downsizing \citep{Juneau2005}.  In contrast, the SFRs per stellar mass or specific SFRs (sSFR) are high.  Local LCBGs have sSFRs ranging from 0.065~$-$~1.7~Gyr$^{-1}$, characterizing them as star forming but not starbursting \citep{Luo2014}.  We find that this sample of local LCBGs has high metallicities, on average exceeding solar, as expected for galaxies thought to have experienced substantial star formation in the past.  Note that the metallicities assume a solar value of 12+log(O/H)~=~8.69 \citep{Tremonti2004}.

Table~\ref{table:properties} also includes R$_e(r)$, the source Petrosian half-light or effective radius in the \textit{r}~band, retrieved from the SDSS~DR7 and converted to kpc using each galaxy's distance.  The effective radii range from 0.60 to 3.5~kpc; the mean is 1.7~kpc.

\subsubsection{Environment}
\label{environment}
\begin{figure}[ht]
\epsscale{1}
\plotone{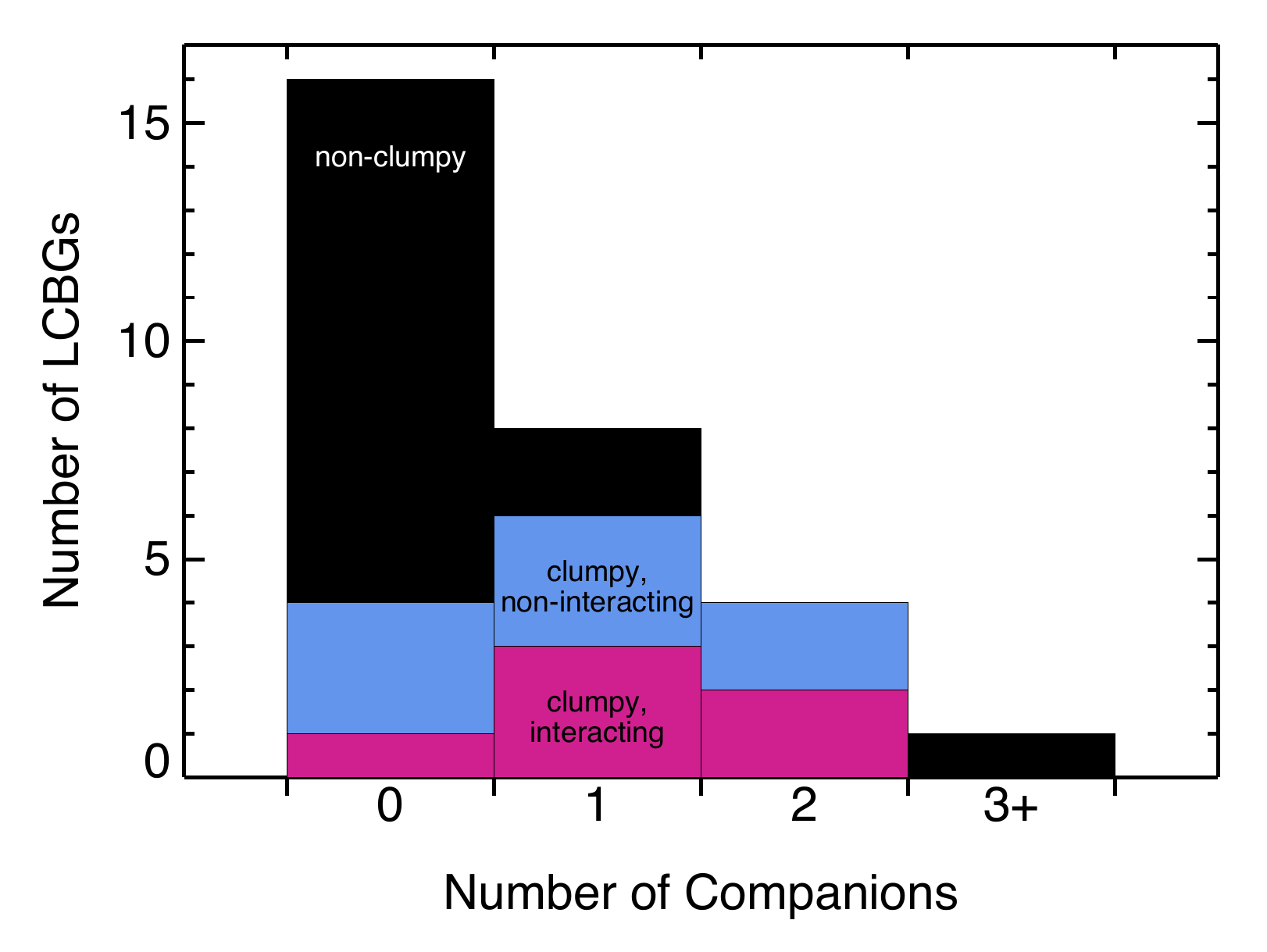}
\caption{The number of local LCBGs with zero, one, two, or more companions.  Approximately half (45$\pm$12\%) of our local LCBGs have at least one companion.  While LCBGs with and without companions show clumpy morphologies, it appears more common for clumpy galaxies, interacting or not, to have companions.  Note the Poisson errors due to our small sample size.}
\label{fig:companions}
\end{figure}

We studied both the local and global environments of local LCBGs to explore possible star formation triggering mechanisms.  As shown in Figure~\ref{fig:companions}, and in line with previous work \citep{LCBGI, Perez2011}, nearly half of the 29 local LCBGs have one or more optical companions within 100~kpc and 1000~km~s$^{-1}$ catalogued in the NASA/IPAC Extragalactic Database (NED)\footnote{http://ned.ipac.caltech.edu/}.  Figure \ref{fig:companions} also illustrates that it is more common for LCBGs with companions to be clumpy as compared to isolated LCBGs.

We note the small numbers in our sample which lead to large uncertainties in such statements. For example, 20\% of the non-clumpy galaxies have companions.  Since this statistic is based on three galaxies, out of a total of 15 non-clumpy galaxies, the standard deviation on the stated 20\% is 12\%, based on Poisson statistics.

Using galaxy morphology (i.e. the visual appearance of tidal tails or other non-regular features) and information available in NED, six local LCBGs appear to be interacting or merging systems. Such galaxies are indicated with an I in Figures~\ref{fig:LCBG_panel_not_clumpy} and \ref{fig:LCBG_panel_clumpy}.  The interactive nature of some of these six sources is not obvious from the small images, centered on the LCBGs, presented in these figures.  Given the sensitivity of SDSS (see Section \ref{sample}) and Haynes' (2008) finding that there are very few optically-dark \hi~rich galaxies, we are confident we are not missing a significant number of LCBG companions.  

\begin{figure}[ht]
\epsscale{1}
\plotone{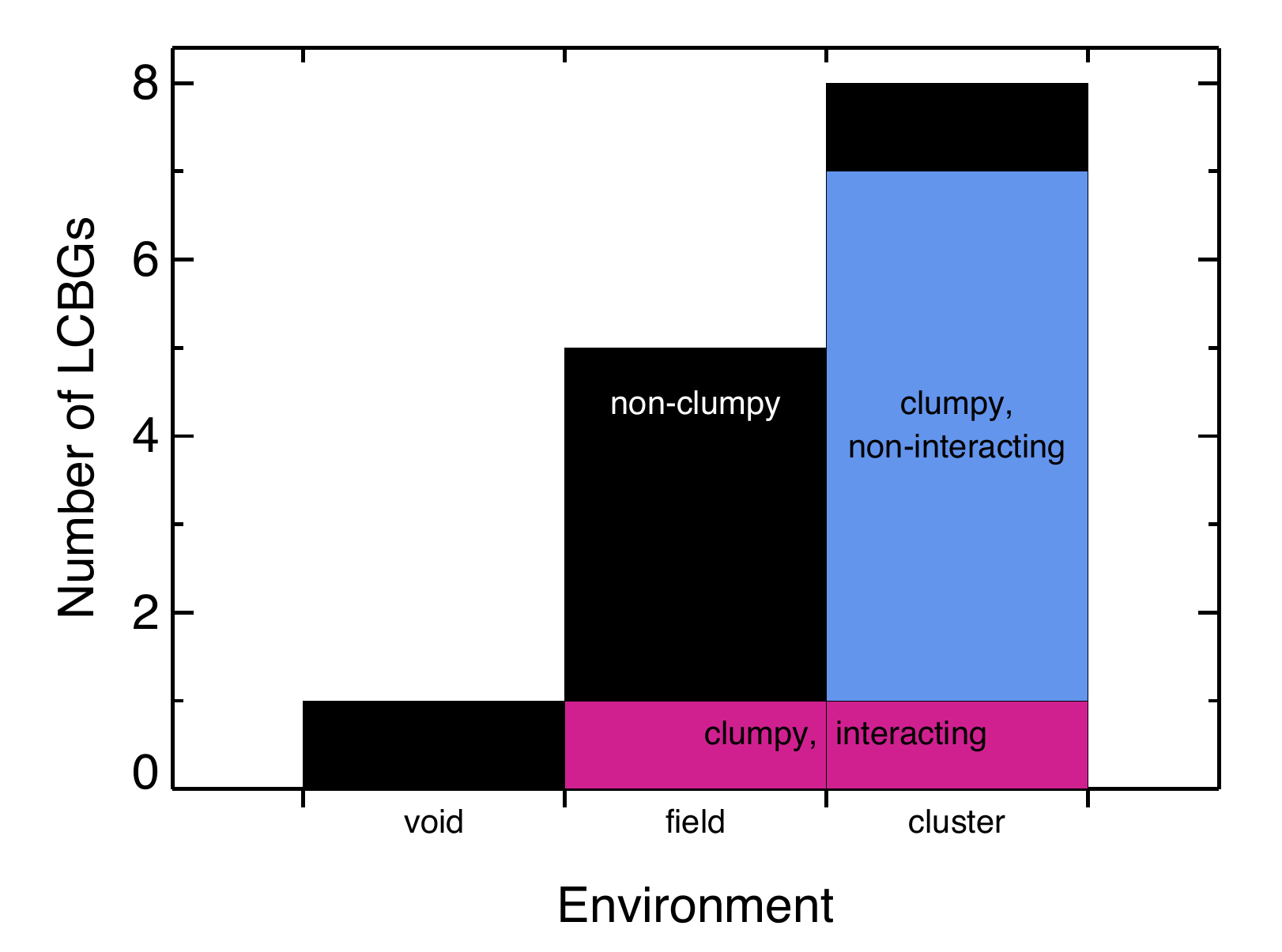}
\caption{The environment of 14 local LCBGs.  Approximately 60\% (57$\pm$20\%) of our local sample of LCBGs are found in clusters, $\sim$40\% (36$\pm$16\%) in the field and only one is located in a void.  Of the eight galaxies in a cluster environment, seven are clumpy with only one showing signs of an interaction.  Only one of the five field galaxies is clumpy, and it shows signs of an interaction.  Note the Poisson errors due to our small sample size.}
\label{fig:density}
\end{figure}

We examined the large scale environment of local LCBGs by finding their 
locations within a density field reconstructed from the three 
dimensional galaxy distribution.  We applied the Delaunay Tessellation 
Field Estimator \citep{Schaap2000, vandeWeygaert2009, Cautun2011} to the sixth SDSS data release galaxy sample, 
following \citet{Platen2009} and \citet{Kreckel2011}, and calculated the 
environmental density contrast for 14 galaxies in our sample.  
As shown in Figure \ref{fig:density}, these 14 LCBGs span the range of environmental densities, from voids to clusters, but only one galaxy was found in a void. Seven of the eight cluster LCBGs are clumpy and only one of these is an interacting system.  In comparison, only one of the five field LCBGs is clumpy, and it is an interacting system.

Note that we were only able to calculate the environmental density contrast for 14 of the 29 galaxies. The remaining 15 galaxies were excluded either because they fall outside the central SDSS redshift survey entirely, or are too near the edge of the survey to have well constrained environmental density measurements.  However, this sub-sample of 14 sources is representative of the parent sample:  The averages of the measured properties (M$_\star$, M$_{HI}$, sSFR, f$_{HI}$, and metallicity) and the selection criteria in the subsample fall within one standard deviation of the average properties of the larger sample.  A Kolmogorov-Smirnov test was also performed to compare the distribution of M$_\star$ in the two samples.  The resulting p-value, 0.836, indicates an 83.6\% likelihood that the two samples are drawn from the same distribution.

\subsection{\hi~Properties}
We obtained original single-dish \hi~observations for 13 LCBGs, using Arecibo, the Green Bank Telescope and the Nan\c{c}ay Decimetric Radio Telescope.  We used archival data from \citet{Springob2005}'s digital \hi~archive for another 13 sources.  Finally, archival data of two sources in the \hi~Parkes All Sky Survey (HIPASS, \citealt{HIPASS}) and one in the Arecibo Legacy Fast Arecibo L-Band Feed Array Survey (ALFALFA, \citealt{ALFALFA}) were used.  The acquisition, calibration and reduction of our original observations and the calculation of \hi~masses ($M_{HI}$), from original and archival data, will be detailed in future work.  We expect that the molecular gas content of these galaxies is low, $\sim$10\%, based on carbon monoxide observations of a different sample of 20 local LCBGs selected in the same way \citep{LCBGII}. In addition, observational and theoretical results suggest that the ratio of molecular to atomic hydrogen both decreases with mass and redshift \citep{Bothwell2009, Catinella2010, Lagos2011} so that in our local, low mass galaxies the atomic gas should be dominant. 

\begin{figure}[ht]
\epsscale{1}
\plotone{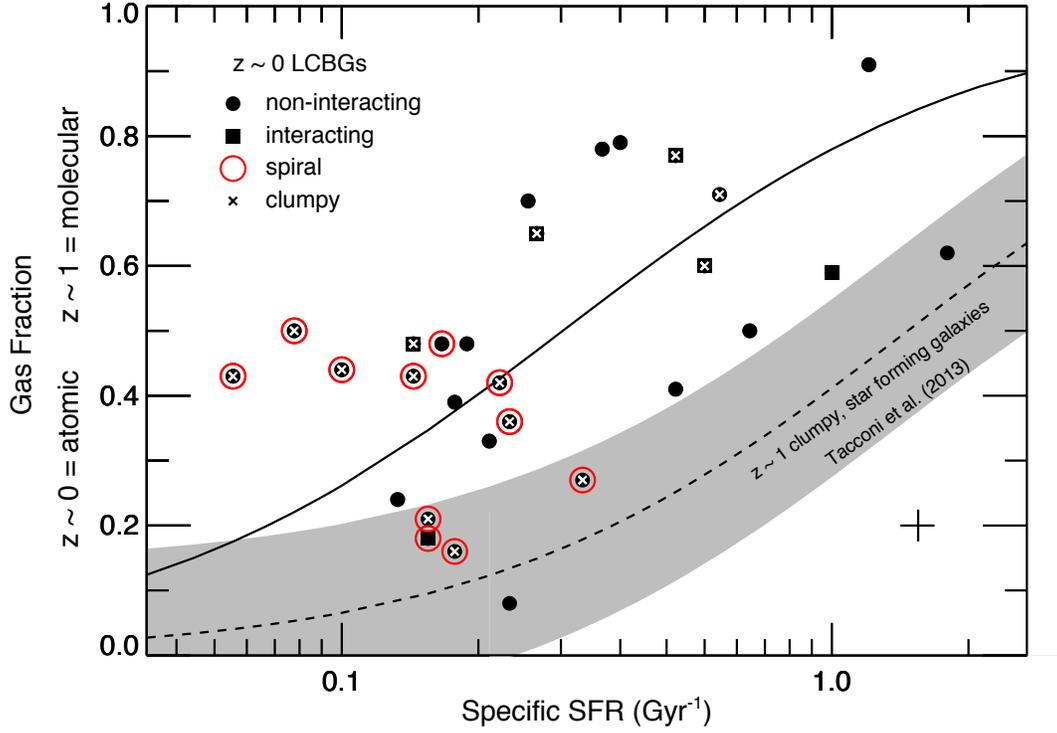}
\caption{The atomic gas fraction versus specific star formation rate for local LCBGs.  The average 1~$\sigma$ error on the data points is shown in the lower right corner.   
The solid line shows the best fit to Equation (1) yielding a constant $\tau_g$ of 3.59~$\pm$~0.10~Gyr. We compare local LCBGs to the z~=~1$-$1.5 clumpy, star forming galaxies studied by \citet{Tacconi2013}.  The dashed line shows the best fit to these galaxies by Equation (1) which yields a
$\tau_g\sim$0.7 Gyr.  The gray shaded region indicates their standard deviation, 0.14.}
\label{fig:fg_ssfr}
\end{figure}

We find that local LCBGs have high atomic gas fractions, defined as $f_{HI}~=~M_{HI}~(M_{HI} + M_\star)^{-1}$, averaging 48\% but ranging from 8\%$-$91\% (Table~\ref{table:properties}).  We considered whether some galaxies might have overestimated values of M$_{HI}$ due to the presence of companions in the relatively large ($\sim$4$-$15$'$) single-dish beams.  (Nearly half of our sample has at least one optical companion, as discussed in Section \ref{environment}.)  However, if only the LCBGs without companions are considered, the average and range of the gas fractions are nearly the same (49\%, 8\%$-$79\%).  We find that the sSFRs of local LCBGs increase with gas fraction (Figure \ref{fig:fg_ssfr}), indicating that f$_g$ is responsible for enhanced SFRs.

The gas fraction and sSFR are related through the gas depletion timescale, $\tau_g~=~M_{HI}~SFR^{-1}$, as shown in the following equation:
\begin{equation}
f_{HI}~=~\frac{1} {1 + (sSFR \times \tau_g)^{-1} } 
\end{equation}
We find, as shown by the solid line in Figure \ref{fig:fg_ssfr}, that local LCBGs are fit to Equation~(1) with a constant $\tau_g$~=~3.59~$\pm$~0.10~Gyr.  The standard deviation between the measured gas fractions and the gas fractions calculated using this fit to Equation~(1) is $\pm$0.18.  \citet{Tacconi2013} found that clumpy, star forming galaxies at z~$\sim$~1 when fit to this relation had a shorter $\tau_g$ ($\sim$~0.7~Gyr) with a similar standard deviation (dotted line and shaded regions in Figure \ref{fig:fg_ssfr}).  The local gas fractions are atomic, while those measured by \citet{Tacconi2013} are molecular.  Although we cannot directly compare these two findings, there is likely additional atomic gas in the high-redshift galaxies, raising their gas fractions into the range we observe.  (Direct measurements of atomic hydrogen are not available for high redshift sources.)  Both populations show a trend of increasing sSFRs with increasing gas fraction.

\section{DISCUSSION}
Our z~$\sim$~0 sample of LCBGs is similar to high redshift SFGs in terms of morphology, including clumpiness; effective radii; clump sizes; specific star formation rates and gas fractions.  The size of clumps in local LCBGs, $\sim$1~kpc, is similar to that observed in clumpy galaxies at higher redshifts \citep{Elmegreen2004, Genzel2008, Forster2009}.  The effective radii of local LCBGs are similar to the lower end of the range found in samples of higher redshift SFGs.  In our local sample of LCBGs, R$_e(r)$ ranges from 0.60 to 3.5 kpc (mean~=~1.7 kpc), while, for example, the SFGs studied by \cite{Tacconi2013} at z~$\sim$~1$-$3 have effective radii of 1.0 to 11.2~kpc (mean~=~4.4~kpc).  \cite{Forster2009} and \cite{Epinat2012} report similar ranges and mean effective radii for SFG samples at similar redshifts.  Both \cite{Forster2009} and \cite{Epinat2012} note that those sources they identify as non-rotators are more compact than those that are rotationally supported.  For example, \cite{Epinat2012} find a mean effective radius of 3.8~kpc for rotating SFGs and 2.7~kpc for non-rotating SFGs at high-z.  Note that the effective radii are measured in a variety of ways and at a variety of wavelengths.

The sSFRs of local LCBGs are also similar to the low end of the range found in star forming, clumpy galaxies at z~$\sim$~1$-$3 \citep{Bournaud2012,Tacconi2013}, and, as at higher redshift (e.g. \citealt{Tacconi2013}), we find that the sSFRs of local LCBGs increase with gas fraction (Figure \ref{fig:fg_ssfr}).
The atomic gas fractions of local LCBGs are much higher than typical, local spiral galaxies \citep{Schombert2001} but similar to the molecular gas fractions of high redshift clumpy SFGs. For example, such galaxies have average molecular gas fractions of $\sim$20$-$30\% at z~$\sim$~0.1 \citep{Fisher2014} and $\sim$30$-$50\% at z~$\sim$~1$-$3 \citep{Daddi2010, Tacconi2013}.   

We now return to the question of the cause of the enhanced SFRs and sometimes clumpy morphologies in local LCBGs.  Our local sample seems to be divided into three groups: interacting galaxies (e.g. SDSS J082604.32+455803.4); clumpy, spiral galaxies (e.g. SDSS J231433.12+001409.5); and smooth, non-spiral galaxies with regular shapes (e.g. SDSS J081214.88+350925.6).  This division can be seen in Figure~\ref{fig:fg_ssfr} where galaxies are coded according to appearance.

Only six of 29 local LCBGs have irregular shapes that suggest they are experiencing strong galaxy-galaxy or galaxy-cluster interactions.  Mergers and major galaxy-galaxy interactions cause a variety of galaxy morphologies, with concentrated or extended star formation and clumpy or non-clumpy appearances, depending on the characteristics of the two galaxies (e.g. \citealt{Barnes2004, Mandelker2013}).  Interactions between a galaxy and a cluster potential or the intracluster medium can cause similar morphologies and star formation properties, including the enhancement of clump formation via ram compression (e.g. \citealt{Tonnesen2011}). While some local LCBGs appear to be undergoing star formation induced by major interactions, or perhaps by cluster interaction, their small numbers suggest neither is a dominant star formation trigger at this epoch.

The remaining 23 galaxies separate into two roughly equal groups: 1. clumpy galaxies with spiral morphologies and 2. smooth, non-spiral galaxies with regular shapes.  We refer to these as the clumpy and non-clumpy groups, respectively. (We exclude galaxies undergoing major interactions in the following discussion.) 
While both clumpy and non-clumpy galaxies have gained enough gas to trigger their current star formation episodes, different mechanisms may be at work, explaining the differing characteristics of the two groups.

The clumpy galaxies, in addition to tending towards spiral morphologies, also tend to be redder (B$-$V~$\gtrsim$~0.4) and to have larger effective radii (R$_{e}$($r$)~$\gtrsim$~1.5~kpc), higher stellar masses (M$_\star$~$\gtrsim$~3$\times$10$^9$~M$_\sun$), lower gas fractions and lower sSFRs (Figure \ref{fig:fg_ssfr}) than the non-clumpy galaxies.  The two groups do not show differences in the ranges of M$_{HI}$, dynamical mass or brightness (Garland et al., in prep).  Clumpy galaxies tend to be found in clusters and have companions, as illustrated in Figures \ref{fig:companions} and \ref{fig:density}.  This suggests that the environment influences the mechanisms by which each group of galaxies has built up high gas fractions.  

Based on their high gas fractions and extended, clumpy star formation, it seems likely that the clumpy galaxies are experiencing violent gravitational instabilities that fragment their disks without destroying them.  Their locations within clusters seem to rule out the possibility of cold flows \citep{Keres2005}.  It seems more likely that these galaxies are accreting gas from companions or from surrounding intracluster medium that has been enriched by stripped satellites.  Such gas inflow, if smooth (i.e. not a sequence of clumpy, minor mergers), may lead to the same type of violent gravitational instabilities and disk fragmentation as cold streams at higher redshifts.  It is also possible that clump formation is enhanced in these galaxies due to ram compression (e.g. \citealt{Tonnesen2011}).  Certainly, whatever the mode of gas accretion and clump formation, it is not destroying the disks in these galaxies, which makes major interactions unlikely as an explanation for their characteristics.  (While models and simulations of major mergers of gas rich galaxies suggest disks may not be destroyed or may reform quickly (e.g. \citealt{Hopkins2009}), the resulting galaxy would likely include a highly concentrated central starburst, not smaller clumps of star formation spread throughout, as we observe.)

Non-clumpy local LCBGs are not in cluster environments and have small enough stellar masses that simulations place these galaxies in the regime of cold accretion, even at their low redshifts \citep{Keres2005}. While these galaxies do not exhibit the clumpy signatures of violent disk instabilities driven by such streams, the conditions that led to their apparent lack of clumps are unclear.  Our observations cannot distinguish between scenarios where clumps never formed; are present but difficult to detect due to the smaller effective radii ($\lesssim$1.5~kpc) of non-clumpy galaxies; or formed in the past but have since migrated to the galaxy centers, possibly destroying any spiral arms present in the process.  One way for LCBGs to accrete gas via cold streams without forming clumps is if the cold streams are themselves clumpy. Simulations by \citet{Dekel2009} suggest that clumpy streams disrupt disks and prevent clump formation.  It is also possible that a series of minor mergers would have the same effect, or that these LCBGs experienced major interactions in the past but the signatures are no longer visible.  Neutral hydrogen mapping, optical integral field spectroscopy, and higher resolution optical images of local LCBGs would help to disentangle these multiple formation scenarios.  

\section{CONCLUSIONS}
1. Local LCBGs appear similar to high redshift SFGs.  Both populations have similar morphologies and clump sizes.    The sSFRs and effective radii of local LCBGs are similar to the low ends of the ranges found in higher redshift SFGs. The total gas fractions in local LCBGs are as high as some of the common clumpy SFGs at z~$\sim$~1$-$3.  However, local LCBGs do have lower stellar masses and proportions of atomic to molecular gas, as expected due to cosmic downsizing.  These objects offer a local laboratory for investigation of star formation in gas-rich galaxies.

2. We find three distinct types of local LCBGs:
\begin{itemize}
\item Those with star formation likely triggered by strong galaxy-galaxy or galaxy-cluster interactions ($\sim$20\%).
\item Clumpy, spiral galaxies with star formation likely triggered by smooth accretion ($\sim$40\%).  As these galaxies lie primarily in clusters and have companions, this gas likely originates from tidally disrupted companions or intracluster medium polluted by stripped satellites, rather than cold streams.
\item Non-clumpy, non-spiral field galaxies with centrally concentrated morphologies and smaller effective radii and stellar masses ($\sim$40\%).  Possibilities for their high gas fractions and specific SFRs include clumpy accretion (clumpy cold streams or minor mergers), smooth cold streams, or major interactions with no remaining visible signatures.
\end{itemize}

3. The dichotomy in characteristics, including morphologies, of local LCBGs is a result of different local and global environments which affect the method of building large gas fractions in these galaxies.
 
\acknowledgments
CAG was partly supported for this work by Castleton State College through an Advanced Study Grant.  DJP and KR acknowledge partial support from NSF CAREER grant AST-1149491.  M-MML was partly supported by NSF grant AST11-09395 and the Alexander von Humboldt-Stiftung.  

The Arecibo Observatory is sponsored by the National Science Foundation.

The Nan\c{c}ay Decimetric Radio Telescope is funded by the Observatoire de Paris, the CNRS (Institut National des Sciences de l'Univers-INSU of the Centre National de la Recherche Scientifique), and the Conseil G\'{e}n\'{e}ral of the D\'{e}partement du Cher.

The National Radio Astronomy Observatory is a facility of the National Science Foundation operated under cooperative agreement by Associated Universities, Inc.

The Parkes Telescope is part of the Australia Telescope which is funded by the Commonwealth of Australia for operation as a National Facility managed by CSIRO.

This research has made use of NASA's Astrophysics Data System and the NASA/IPAC Extragalactic Database (NED), which is operated by the Jet Propulsion Laboratory, California Institute of Technology, under contract with the National Aeronautics and Space Administration.

Funding for the SDSS has been provided by the Alfred P. Sloan Foundation, the Participating Institutions, the National Science Foundation, the U.S. Department of Energy, the National Aeronautics and Space Administration, the Japanese Monbukagakusho, the Max Planck Society, and the Higher Education Funding Council for England. The SDSS Web Site is http://www.sdss.org/.

The SDSS is managed by the Astrophysical Research Consortium for the Participating Institutions. The Participating Institutions are the American Museum of Natural History, Astrophysical Institute Potsdam, University of Basel, University of Cambridge, Case Western Reserve University, University of Chicago, Drexel University, Fermilab, the Institute for Advanced Study, the Japan Participation Group, Johns Hopkins University, the Joint Institute for Nuclear Astrophysics, the Kavli Institute for Particle Astrophysics and Cosmology, the Korean Scientist Group, the Chinese Academy of Sciences (LAMOST), Los Alamos National Laboratory, the Max-Planck-Institute for Astronomy (MPIA), the Max-Planck-Institute for Astrophysics (MPA), New Mexico State University, Ohio State University, University of Pittsburgh, University of Portsmouth, Princeton University, the United States Naval Observatory, and the University of Washington.
\clearpage

\begin{table}
\caption{Properties of D~$\leq$~76~Mpc LCBG Sub-Sample}
\label{properties}
\begin{tabular}{lll}
\hline
Property & Mean & Range \\
\hline
Distance & 55 Mpc & 20 $-$ 76 Mpc\\
$R_e$$(r)$ & 1.7 kpc & 0.60 $-$ 3.5 kpc \\
M$_\star$ & 3.2$\times$10$^9$ M$_\odot$ & 5.4$\times$10$^8$ $-$ 7.4$\times$10$^9$ M$_\odot$ \\
SFR & 0.88 M$_\odot$ yr$^{-1}$ & 0.22 $-$ 3.98 M$_\odot$ yr$^{-1}$\\
sSFR & 0.37 Gyr$^{-1}$ & 0.065 $-$ 1.7 Gyr$^{-1}$\\
metallicity & 150\% solar  & 81 $-$ 300\% solar\\
$f_{HI}$ & 48\% & 8 $-$ 91\% \\
\end{tabular}
\label{table:properties}
\end{table}

\clearpage

\clearpage


\end{document}